\documentclass[11pt]{article}

\usepackage{amsfonts}
\usepackage{epsfig}
\usepackage{latexsym}
\usepackage{amsmath}
\usepackage{amssymb}
\usepackage{float}

\def\bequ{\begin{equation}}
\def\eequ{\end{equation}}
\def\barr{\begin{array}}
\def\earr{\end{array}}

\def\ben{\begin{equation}}
\def\een{\end{equation}}
\def\bena{\begin{eqnarray}}
\def\eena{\end{eqnarray}}

\def\spa#1{\phantom{\fbox{\rule[-#1cm]{0cm}{0cm}}}}

\def\b1{e^0}

\newcommand{\be}{\begin{equation}}
\newcommand{\ee}{\end{equation}}
\def\bea{\begin{eqnarray}}
\def\eea{\end{eqnarray}}

\def\be{\begin{equation}}
\def\ee{\end{equation}}
\def\bea{\begin{eqnarray}}
\def\eea{\end{eqnarray}}

\def\lesssim{\mathrel{\hbox{\rlap{\hbox{\lower4pt\hbox{$\sim$}}}\hbox{$<$}}}}
\def\gtrsim{\mathrel{\hbox{\rlap{\hbox{\lower4pt\hbox{$\sim$}}}\hbox{$>$}}}}

\begin{document}
\title{{\LARGE \bf{On the backreaction of frame dragging}}}
\author{
Carlos A. R. Herdeiro$^{1,}$\footnote{crherdei@fc.up.pt}~, \
Carmen Rebelo$^{1,}$\footnote{mrebelo@fc.up.pt}~  and 
Claude M. Warnick$^{2,}$\footnote{cmw50@cam.ac.uk}
\\
\\ $^{1}${\em Departamento de F\'\i sica e Centro de F\'\i sica do Porto}
\\ {\em Faculdade de Ci\^encias da Universidade do Porto}
\\ {\em Rua do Campo Alegre, 687,  4169-007 Porto, Portugal}
\\
\\ {$^{2}${\em D.A.M.T.P.}}
\\ {\em  Cambridge University}
\\ {\em Wilberforce Road, Cambridge CB3 0WA, U.K.}}

\date{July 2009}       
 \maketitle

\begin{abstract}
The backreaction on black holes due to dragging heavy, rather than test, objects is discussed. As a case study, a regular black Saturn system where the central black hole has vanishing intrinsic angular momentum, $J^{BH}=0$, is considered. It is shown that there is a correlation between the sign of two response functions. One is interpreted as a moment of inertia of the black ring in the black Saturn system. The other measures the variation of the black ring horizon angular velocity with the central black hole mass, for fixed ring mass and angular momentum. The two different phases defined by these response functions collapse, for small central black hole mass, to the thin and fat ring phases. In the fat phase, the zero area limit of the black Saturn ring has reduced spin $j^2>1$, which is related to the behaviour of the ring angular velocity. Using the `gravitomagnetic clock effect', for which a universality property is exhibited, it is shown that frame dragging measured by an asymptotic observer decreases, in both phases, when the central black hole mass increases, for fixed ring mass and angular momentum. A close parallelism between the results for the fat phase and those obtained recently for the double Kerr solution is drawn, considering also a regular black Saturn system with $J^{BH}\neq 0$.
\end{abstract}


\section{Introduction}
Frame dragging in general relativity is usually presented by considering the effect that a dragging source, typically a rotating shell \cite{LT} or a rotating black hole \cite{Wilkins:1972rs}, has on test objects. In this paper we shall consider the following question: ``How does dragging a heavy, rather than a test, object backreact on the dragging source?"

\begin{figure}[h!]
\centering\includegraphics[height=2.7in,width=3.7in]{{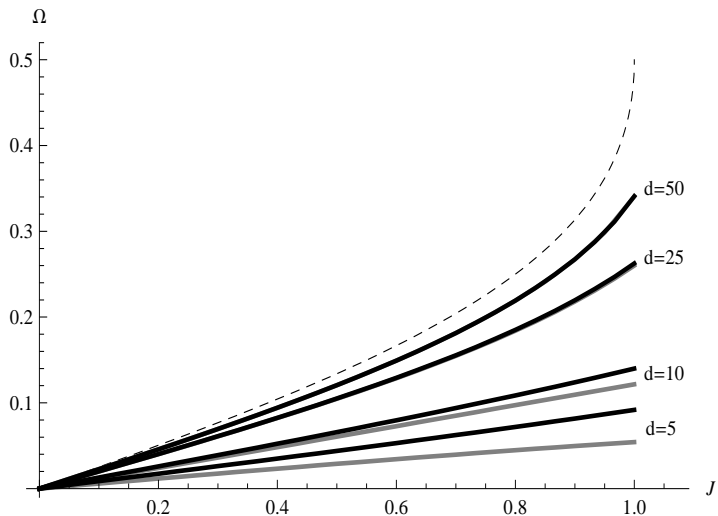}} 
\begin{picture}(0,0)(0,0)
\end{picture}
\caption{Horizon angular velocity as function of the angular momentum, for fixed mass $M=1$, in the double Kerr system. The dashed line corresponds to the infinite distance limit (i.e the usual Kerr solution). The black (grey) lines correspond to the co-rotating (counter-rotating) double Kerr system for different values of the physical distance $d$. Note that in the co-rotating (counter-rotating) case, $\Omega=\Omega_1=\Omega_2$ ($\Omega=\Omega_1=-\Omega_2$).}
\label{doublekerr}
\end{figure}

A hint, as well as a further motivation, to answer this question comes from recent studies of the double Kerr solution in four dimensions \cite{Herdeiro:2008kq,Costa:2009wj}. These studies focused on two special and treatable cases of the general double Kerr \cite{KN}, dubbed counter-rotating and co-rotating solutions. Both are three parameter families of asymptotically flat solutions, obeying the ``axis condition"\footnote{In terms of the rod structure of this solution, this condition amounts to the requirement  that the azimuthal Killing vector field is the eigenvector associated to the finite spacelike rod in between the two black hole horizons.} and describe two equal mass black holes (mass $M$) with the same (co-rotating) or opposite (counter-rotating) angular momentum ($J=J_1=\pm J_2$), at a physical (i.e. proper) distance $d$ from one another. The analysis in \cite{Herdeiro:2008kq,Costa:2009wj} revealed that, in both the counter-rotating and the co-rotating case, the angular velocity of the horizon, $\Omega$, for each black hole, decreases when the two black holes are approached, keeping fixed the mass and the angular momentum of the black holes, computed as Komar integrals - Fig. \ref{doublekerr}. Moreover, it shows that the angular velocity is smaller, at the same physical distance, in the counter-rotating than in the co-rotating case.

 The behaviour of the angular velocity shown in Fig. \ref{doublekerr} is intuitive in the counter-rotating case, since the two black holes are counter-dragging one another. But it is not so in the co-rotating case. In \cite{Costa:2009wj}  it was suggested that the result exhibited in Fig. \ref{doublekerr} reflects two effects for each black hole: 
 \begin{description}
 \item[i)] the co/counter-dragging effect by the other black hole - this is the usual effect one associates to rotational dragging and which can be seen at the level of test objects;
 \item[ii)] the \textit{backreaction} of dragging the other black hole - this is the novel effect we shall discuss, which can be seen in exact multi-black hole solutions but not at the level of test objects.
 \end{description} 
  
  To further the discussion, it proves useful to consider the following response function
 \bequ
 \frac{1}{I(M,J,d)}=\left(\frac{\partial \Omega}{\partial J}\right)_{M,d} \ . \label{mi}\eequ
 The quantity $I(M,J,d)$ is naturally interpreted as the moment of inertia of each black hole in this system. Even though black holes are not rigid objects, their angular velocity is expected to be constant on each connected component of the event horizon, in a stationary configuration.

 Fig. \ref{doublekerr} shows that such moment of inertia is positive. Moreover it shows that the moment of inertia increases as the distance $d$ decreases, for fixed $M$ and $J$, naturally leading to a smaller angular velocity. Our goal is to provide evidence for a correlation between effect ii) and the behaviour of the moment of inertia. With this goal in mind let us consider the following type of systems.

  Consider two black holes in equilibrium in a stationary, axi-symmetric and asymptotically flat spacetime. The first black hole has a fixed
mass $M_1$ and a fixed non-vanishing intrinsic angular momentum $J_1$. The
second black hole has mass $M_2$, which we shall vary. To simplify we take the second black hole to have no intrinsic angular momentum $J_2=0$. In this way, the second black hole will not exert effect i) upon the first black hole, at least to first order.  The variation of the angular velocity of the former black hole $\Omega_1$ with the mass of the latter, 
\bequ
\left(\frac{\partial \Omega_1}{\partial M_2}\right)_{M_1,J_1}  \ , 
\label{introeq} \eequ
gives a well defined notion to decide if there is a slow down effect associated to the dragging of a heavy object and therefore a measure of effect ii). A negative value for \eqref{introeq} shows that the backreaction of dragging black hole two is to slow down black hole one. 

In principle the analysis of \eqref{introeq} could be done in the double Kerr system. However, in order to be in equilibrium, the solution has a strut in between the two black holes, corresponding to a conical singularity. This introduces extra physical parameters and makes the system less ``clean" than desirable. Fortunately, a regular system with the exact properties described in the
previous paragraphs exists, albeit in five rather than four spacetime dimensions. It is the black Saturn solution of the five dimensional
vacuum Einstein equations \cite{Elvang:2007rd}. Let us take the first
black hole to be the black ring and the second to be the central black
hole (having a topologically spherical horizon). Then \eqref{introeq}
expresses how the angular velocity of the black ring
varies, for fixed ring mass and intrinsic angular momentum, when the
mass of the central black hole, with no intrinsic spin, varies. In
particular, if the central black hole mass is small as compared
to the ADM mass of the system, \eqref{introeq} tells us how the angular velocity of
an isolated black ring varies if it has to drag a small central black
hole along. 

\begin{figure}[h!]
\centering\includegraphics[height=1in]{{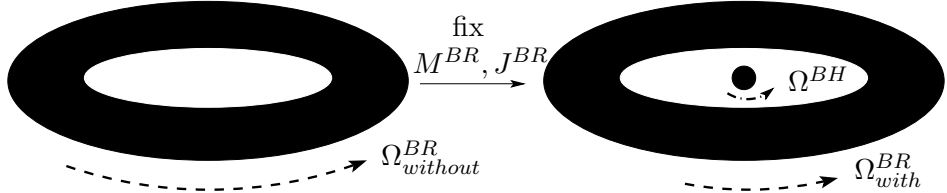}} 
\begin{picture}(0,0)(0,0)
\put(-63,38){$\Omega^{BH}$}
\put(-217,10){$\Omega^{BR}_{without}$}
\put(-38,5){$\Omega^{BR}_{with}$}
\put(-205,45){$M^{BR},J^{BR}$}
\put(-190,59){\rm fix}
\end{picture}
\caption{Illustration of a fat black ring (left) to which a small central black hole without intrinsic angular momentum is added (right), for fixed black ring mass and angular momentum. Effect i) is that $\Omega^{BH}\neq 0$; effect ii) is that $\Omega^{BR}_{without}>\Omega^{BR}_{with}$.}
\label{fatring}
\end{figure}

The first result we shall present here is that the dynamics of effect ii) can actually be richer than expected from the double Kerr example. In the black Saturn there are two phases, corresponding to two different signs for  \eqref{introeq}. Moreover, these two different phases can be correlated to different signs for the moment of inertia of the black Saturn ring:
\bequ 
 \frac{1}{I_{BR}(M^{BH},M^{BR},J^{BR})}=
\left(\frac{\partial \Omega^{BR}}{\partial J^{BR}}\right)_{M^{BR},M^{BH}} \ . \label{d1} \eequ
In the limit of small central black hole mass, the sign of \eqref{introeq} is negative (positive) if the black ring is \textit{fat}
(\textit{thin}). Thus, we shall call the phase in which  \eqref{introeq} is negative (positive) the \textit{fat}
(\textit{thin}) phase. Fat black rings are more akin to singly spinning
Myers-Perry black holes (or Kerr black holes) than thin ones. In particular, when $M^{BH}=0$ ,  \eqref{d1} is positive  for fat black rings, just like the analogous quantity for Myers Perry black
holes or, more generally, for standard physical systems, but it is negative for thin black rings. Thus, we shall interpret our result for the fat phase as evidence that in the (cases described of the) double Kerr system, which have a positive moment of inertia, effect ii) leads to the slow down of the horizon angular velocity  - Fig. \ref{fatring}. 

In the case of thin rings, or more generally in the thin phase, the negative value of \eqref{introeq} is therefore correlated with an unusual property: a negative value of \eqref{d1} means that thin black rings decrease their angular velocity when their angular momentum is increased, at fixed mass. Nevertheless, defining a certain measure of frame dragging, we will show that even in the thin phase the dragging effects are smaller, as measured by an asymptotic observer, when the ring has to drag a central black hole. This measure of frame dragging is actually the `gravitomagnetic clock effect' \cite{CM}, a quantity for which we shall describe a universality behaviour. 

Associated to the behaviour of the horizon angular velocity displayed in Fig. \ref{doublekerr} it was found in \cite{Herdeiro:2008kq} and further developed in \cite{Costa:2009wj} that black holes may have $J>M^2$ in the double Kerr system.\footnote{We shall use geometrised units throughout.} It is well known that fat black rings exist only for a limited reduced spin $j^2<1$. However, in complete analogy with the double Kerr case, we shall observe that in the fat phase of the black Saturn system, rings may have $j^2>1$.

This paper is organised as follows. In section 2 we shall describe the analysis of equation \eqref{introeq} for the black Saturn solution. In section 3 we shall correlate that response function with the moment of inertia of the black ring in the black Saturn system \eqref{d1}. In section 4 we discuss the maximum angular momentum that may be carried by the black ring, in the fat phase, for fixed mass, and observe it is larger than that for an isolated black ring with the same mass. In section 5 we shall discuss the `gravitomagnetic clock effect' as a measure of frame dragging. We will exhibit a universality property and apply it to the black Saturn system. In section 6 we shall briefly consider the black Saturn system with a central black hole having non-zero angular momentum, to compare the importance of effects i) and ii) in this system.
In section 7 we shall close with a summary of results and some final remarks.

\section{Black Saturn with $J^{BH}=0$: $(\partial \Omega^{BR}/\partial M^{BH})_{M^{BR},J^{BR}}$}
\begin{figure}[h!]
\centering\includegraphics[height=1.3in,width=4.5in]{{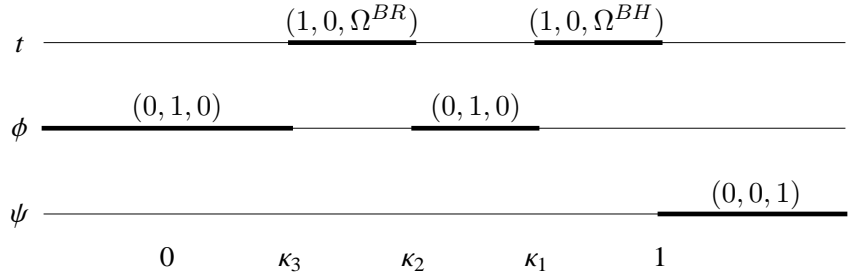}} 
\begin{picture}(0,0)(0,0)
\put(-280,60){${(0,1,0)}$}
\put(-167,60){${(0,1,0)}$}
\put(-61,27){${(0,0,1)}$}
\put(-222,92){${(1,0,\Omega^{BR})}$}
\put(-130,92){${(1,0,\Omega^{BH})}$}
\end{picture}
\caption{Significance of the parameters $\kappa_i$ in terms of the rod structure (including rod directions) of the black Saturn solution. Apparently the solution is determined by five parameters: the three finite rod sizes and the two timelike rod directions. However, the regularity constraint and the $J^{BH}=0$ condition reduce the number of free parameters to three.}
\label{rodstructure}
\end{figure}

The (balanced) black Saturn  \cite{Elvang:2007rd} is a four parameter
family of asymptotically flat solutions of the five dimensional vacuum Einstein
equations. Physically, the four parameters are the central black
hole mass $M^{BH}$ and angular momentum $J^{BH}$  and the ring mass $M^{BR}$ and angular
momentum  $J^{BR}$. All these quantities are computed as Komar integrals. We shall focus on the case with vanishing intrinsic angular momentum for the central black hole
$J^{BH}=0$. Despite simplifying considerably in this special case, the metric is still rather involved. It is explicitly given in \cite{Elvang:2007rd}.  But the physical quantities become quite simple, using the parameterisation therein: 
\bequ
M^{BH}=\frac{3\pi L^ 2}{4}(1-\kappa_1) \ , \qquad M^{BR}=\frac{3\pi L^ 2}{4}\kappa_2 \ , \label{komarmass}\eequ
\bequ
J^{BH}=0 \ , \qquad J^{BR}=\pi L^3\sqrt{\frac{\kappa_2\kappa_3}{2\kappa_1}} \ , \eequ
where $L$ is a length scale. The three dimensionless parameters $\kappa_i$ obey
\bequ
0\le \kappa_3\le \kappa_2<\kappa_1\le 1 \ , \label{order} \eequ
and are not independent in the balanced solution. Rather, the absence of a conical singularity requires 
\bequ
\kappa_1-\kappa_2=\sqrt{\kappa_1(1-\kappa_2)(1-\kappa_3)(\kappa_1-\kappa_3)}
\ . \eequ
The parameters $\kappa_i$ determine the breakpoints in the rod structure - Fig \ref{rodstructure}.
\spa{0.3cm}

Solving for $\kappa_3$, the solution consistent with (\ref{order}) is
\bequ
\kappa_3=\frac{1+\kappa_1}{2}-\sqrt{\left(\frac{1+\kappa_1}{2}\right)^2-\frac{(2-\kappa_1-\kappa_2/\kappa_1)\kappa_2}{1-\kappa_2}}
\ . \eequ
Thus, the physical quantities are now functions of the dimensionless
parameters $\kappa_1,\kappa_2$ and the length scale $L$. 

We shall be interested in the angular velocities of the horizons,
which are
\bequ
\Omega^{BH}=\frac{1}{L}\sqrt{\frac{\kappa_2\kappa_3}{2\kappa_1}} \ ,
\qquad \Omega^{BR}=\frac{1}{L}\sqrt{\frac{\kappa_1\kappa_3}{2\kappa_2}} 
\ . \eequ
The fact that $J^{BH}=0$ and $\Omega^{BH}\neq 0$ manifests the frame dragging due to the angular momentum of the black ring, as observed in \cite{Elvang:2007rd}. This is the usual effect also seen at the level of test objects. In order to see the backreaction of frame dragging we shall analyse 
how the angular velocity of the black ring varies when  $M^{BH}$ varies, keeping fixed $M^{BR}$ and $J^{BR}$. It will be useful, in the following, to consider the \textit{reduced spin} of the black ring, $j$, the \textit{relative mass}, $m$, and the \textit{dimensionless angular velocity} of the black ring horizon, $\omega^{BR}$:
\bequ
j^2\equiv \frac{27\pi}{32}\frac{(J^{BR})^2}{(M^{BR})^3} \ , \qquad m=\frac{M^{BH}}{M^{BR}} \ , \qquad \omega^{BR}=\Omega^{BR}\sqrt{\frac{8}{(3\pi)M^{BR}}} \ .
\eequ
Analogous quantities to $j$ and $\omega^{BR}$ can be defined for the central black hole. These shall be used in section \ref{sectionj}.

Expressing $L=L(M^{BR},\kappa_2)$,
then
\bequ
M^{BH}=\frac{1-\kappa_1}{\kappa_2}M^{BR} \ , \label{1par}
\eequ
\bequ J^{BR}=\left(\frac{4M^{BR}}{3\pi^{1/3}}\right)^{3/2}\frac{1}{\kappa_2}\sqrt{\frac{\kappa_3}{2\kappa_1}} \ , \label{2par}\eequ
and 
\bequ
\Omega^{BR}=\sqrt{\frac{3\pi \kappa_1\kappa_3}{8M^{BR}}}
  \ . \label{3par}\eequ
From \eqref{2par}, requiring $dM^{BR}=0=dJ^{BR}$, it follows that
\bequ
d\kappa_2=-\frac{\kappa_2}{\kappa_1}\left(\frac{\kappa_1\partial \kappa_3/\partial
    \kappa_1-\kappa_3}{\kappa_2\partial
    \kappa_3/\partial \kappa_2-2\kappa_3}\right)d\kappa_1 \ . \eequ
Then, from \eqref{1par} 
\bequ
dM^{BH}=-\frac{M^{BR}}{\kappa_2}\left(\frac{(\kappa_1-1)\partial
  \kappa_3/\partial \kappa_1+\kappa_2\partial \kappa_3/\partial
  \kappa_2+\kappa_3/\kappa_1-3\kappa_3}{\kappa_2\partial
  \kappa_3/\partial \kappa_2-2\kappa_3}\right)d\kappa_1 \ , \eequ
and from \eqref{3par}
\bequ
d\Omega^{BR}=\sqrt{\frac{3\pi\kappa_3}{8M^{BR}\kappa_1}}\left(\frac{-\kappa_1\partial
  \kappa_3/\partial \kappa_1+\kappa_2\partial \kappa_3/\partial
  \kappa_2-\kappa_3}{\kappa_2\partial
  \kappa_3/\partial \kappa_2-2\kappa_3}\right)d\kappa_1 \ . \eequ
Thus
\bequ
\left(\frac{\partial \Omega^{BR}}{\partial M^{BH}}\right)\big|_{M^{BR},J^{BR}}=f(\kappa_1,\kappa_2)\, \sqrt{\frac{3\pi\kappa_3\kappa_2^2}{8(M^{BR})^3\kappa_1}} \ , \eequ
where
\bequ
f(\kappa_1,\kappa_2)=
\frac{\kappa_1\partial
  \kappa_3/\partial \kappa_1-\kappa_2\partial \kappa_3/\partial
  \kappa_2+\kappa_3}{(\kappa_1-1)\partial
  \kappa_3/\partial \kappa_1+\kappa_2\partial \kappa_3/\partial
  \kappa_2+\kappa_3/\kappa_1-3\kappa_3}  \ . \label{f} \eequ

The sign of the function  $f(\kappa_1,\kappa_2)$ determines if the black
ring will increase or decrease its horizon angular velocity when the
mass of the central black hole varies. To compute this sign, let us
assume that the central black hole has a very small mass, i.e. 
\bequ
\kappa_1=1-\epsilon \ , \qquad 0<\epsilon\ll 1 \ . \eequ
Then, a computation gives
\bequ
f(\kappa_1=1-\epsilon,\kappa_2)=-\left(\frac{4}{3}-\frac{\kappa_2}{1-\sqrt{1-\kappa_2}}\right)^{-1}+\mathcal{O}(\epsilon)
\ . \eequ
The function $f(1,\kappa_2)$ is negative for $8/9<\kappa_2<1$ and
positive for $0<\kappa_2<8/9$; therefore, for a small central black hole ($\kappa_1\simeq 1$)
\bequ
\left(\frac{\partial \Omega^{BR}}{\partial M^{BH}}\right)\big|_{M^{BR},J^{BR}} \left\{\begin{array}{l} <0 \ , \qquad {\rm for} \
8/9<\kappa_2<1 \\ >0 \ , 
\qquad  {\rm for} \ 0<\kappa_2<8/9  \end{array} \right.  \ . \label{sign} \eequ
To understand the significance of the turning point $\kappa_2=8/9$,
observe that the reduced spin, for $\kappa_1=1$,  
\bequ
j^2=\frac{1-\sqrt{1-\kappa_2}}{\kappa_2^2} \ , \eequ
takes values
\bequ
j^2(\kappa_2=1)=1 \ , \qquad
j^2\left(\kappa_2=\frac{8}{9}\right)=\frac{27}{32} \ , \qquad
j^2(\kappa_2\rightarrow 0)\rightarrow +\infty \ . \eequ
Thus, the turning point for the sign of \eqref{sign} separates thin
rings ($0<\kappa_2<8/9$) from fat rings $8/9<\kappa_2<1$ (see e.g \cite{Elvang:2006dd}). 

\begin{figure}[h!]
\centering\includegraphics[height=3.5in,width=3.5in]{{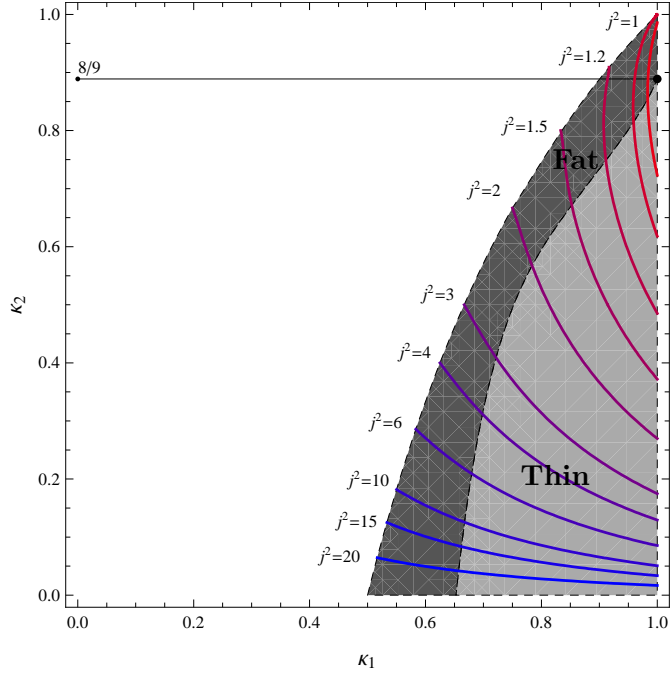}} 
\begin{picture}(0,0)(0,0)
\put(-60,70){{\bf Thin}}
\put(-48,190){{\bf Fat}}
\end{picture}
\caption{Sign of the function $f(\kappa_1,\kappa_2)$ \eqref{f}. The function is negative in the dark grey region (`fat phase') and positive in the light grey region (`thin phase'). The black Saturn solution does not exist outside the coloured region. The lines correspond to trajectories of constant ring mass and angular momentum, i.e only the central black hole mass varies. They have been labelled by the value of the reduced spin $j^2\in [27/32,+\infty [$. These trajectories have the two endpoints on $\kappa_1=1$ for $j^2\le 1$, as it should be from the well known phase diagram of black rings (see e.g \cite{Elvang:2006dd}). As $j^2 \rightarrow  27/32$ the trajectories collapse to the point $(\kappa_1,\kappa_2)=(1,8/9)$.}
\label{sinalf}
\end{figure}

\begin{figure}[h!]
\centering\includegraphics[width=3.2in]{{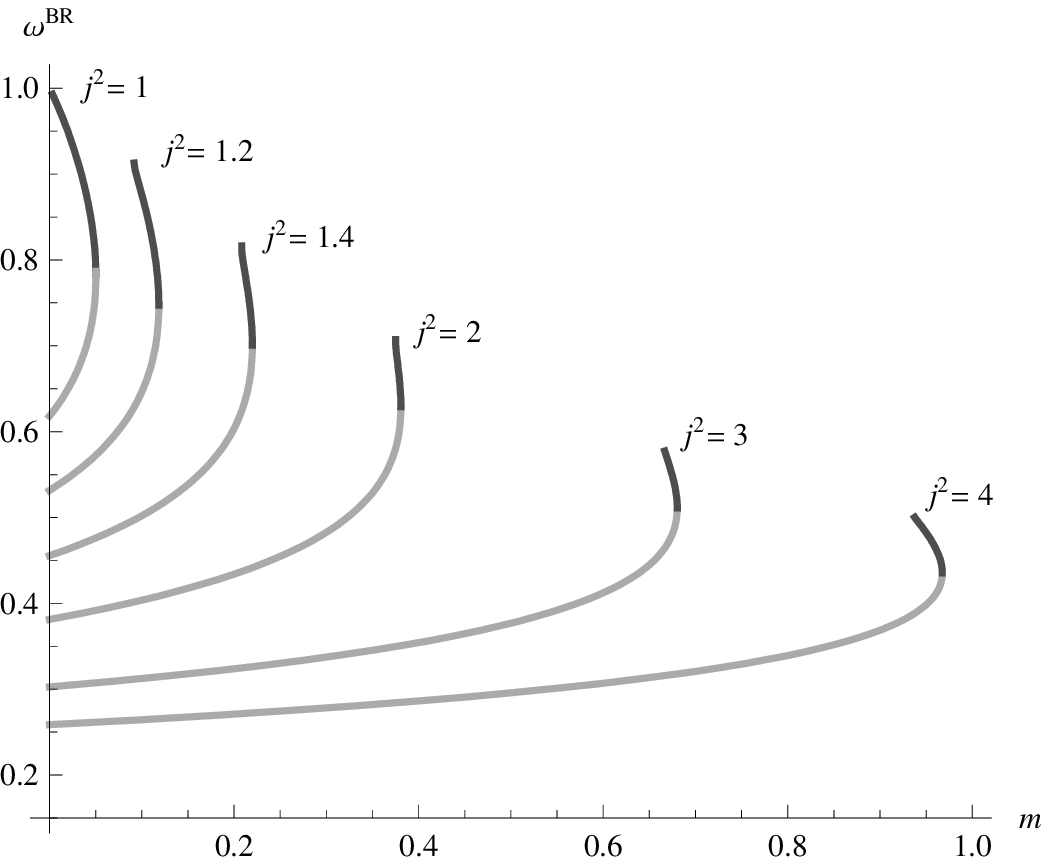}} 
\begin{picture}(0,0)(0,0)
\end{picture}
\caption{Dimensionless angular velocity of the black ring horizon $\omega^{BR}$, in terms of the mass fraction $m$, along trajectories of constant black ring mass and angular momentum. The behaviour in the fat (thin) phase corresponds to the dark (light) grey curves. In the fat (thin) phase, the angular velocity of the ring decreases (increases) as the mass of the black hole increases. Observe that rings in the fat phase may have $j^2>1$, but their angular velocity is always smaller than that of an isolated fat black ring, $\omega^{BR}=1$.}
\label{wvsm}
\end{figure}

Extending the analysis beyond small central black holes yields the result exhibited in Fig. \ref{sinalf}. One finds a region where the response of the ring angular velocity to the increase of the central black hole mass is negative (we call it the \textit{fat phase}) - dark grey in Fig. \ref{sinalf}  - and another region where the response is positive  (\textit{thin phase})  - light grey in Fig. \ref{sinalf}. Along the constant reduced spin trajectories in Fig. \ref{sinalf}, the variation of the ring angular velocity with the black hole mass is shown in Fig. \ref{wvsm}.

Geometrically, the variation of the angular velocity shown in Fig. \ref{wvsm} is accommodated as follows. Computing the proper distance along the equatorial plane,  $l=\int_{\kappa_2}^{\kappa_1}dz \sqrt{g_{zz}}$, between the two black objects shows that, when the central black hole mass increases, the distance increases in the fat phase and decreases in the thin one. This behaviour continues until the two phases meet.

\section{Correlation between two response functions}
\label{correlation}
The result exhibited in the last section for the fat phase supports the interpretation presented in the introduction for the variation of the angular velocity in the double Kerr system: the backreaction on a black object that is dragging a heavy spacetime slows down the black object. But the behaviour of thin phase is the opposite. We will now correlate these results with the behaviour of another response function: the (inverse) moment of inertia.

It is known that thin and fat rings have different responses to a variation of the
angular momentum for fixed mass \cite{Elvang:2006dd}: 
 \bequ
\left(\frac{\partial \Omega^{BR}}{\partial J^{BR}}\right)\big|_{M^{BR}}
  \left\{\begin{array}{l} >0 \ , \qquad {\rm for \ fat \ rings} \ , \
8/9<\kappa_2<1 \\ <0 \ , 
\qquad  {\rm for \ thin \ rings}\ , \ 0<\kappa_2<8/9  \end{array} \right.  \ . \label{sign2} \eequ
The negativity of the response function \eqref{sign2} for thin rings is accommodated geometrically \cite{Elvang:2006dd} as follows. As the angular momentum of a thin ring is increased, at fixed mass, its "radius" increases sufficiently to make a smaller angular velocity consistent with a larger angular momentum.

We shall now observe that the positivity of the response function 
    \bequ
\left(\frac{\partial \Omega^{BR}}{\partial M^{BH}}\right)\big|_{M^{BR},J^{BR}} \ , \label{rf1}
    \eequ
is actually correlated with the negativity of the response function \eqref{d1} in the whole $\kappa_1,\kappa_2$ space, rather than just at $\kappa_1=1$. A straightforward computation shows that
    \bequ
    \left(\frac{\partial \Omega^{BR}}{\partial J^{BR}}\right)\big|_{M^{BR},M^{BH}} 
=\frac{9\pi \kappa_1\kappa_2}{16G(M^{BR})^2}\tilde{f}(\kappa_1.\kappa_2) \ , 
    \eequ
where
\bequ
\tilde{f}(\kappa_1,\kappa_2)=
\frac{(\kappa_1-1)\partial
  \kappa_3/\partial \kappa_1+\kappa_2\partial \kappa_3/\partial
  \kappa_2+\kappa_3-\kappa_3/\kappa_1}{(\kappa_1-1)\partial
  \kappa_3/\partial \kappa_1+\kappa_2\partial \kappa_3/\partial
  \kappa_2+\kappa_3/\kappa_1-3\kappa_3}  \ . \label{ftilde} \eequ

\begin{figure}[h!]
\centering\includegraphics[height=3.5in,width=3.5in]{{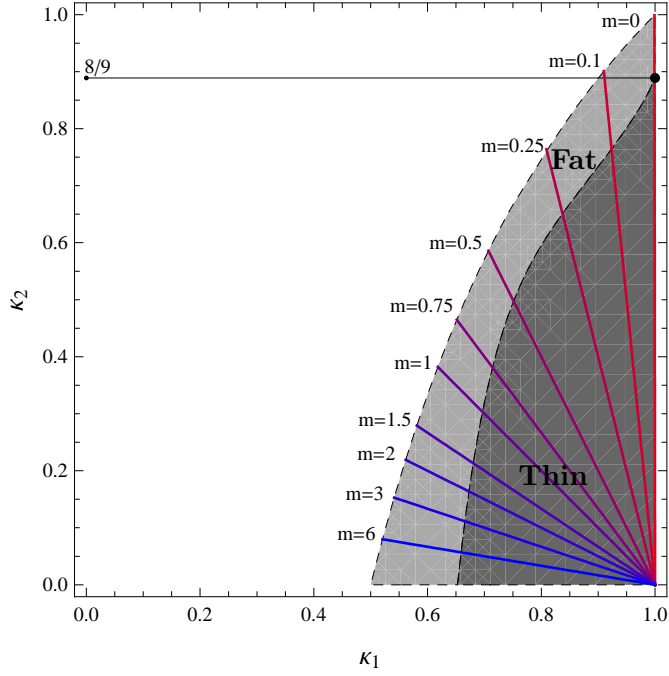}} 
\begin{picture}(0,0)(0,0)
\put(-60,70){{\bf Thin}}
\put(-48,190){{\bf Fat}}
\end{picture}
\caption{Sign of the function $\tilde{f}(\kappa_1,\kappa_2)$ \eqref{ftilde}. The function is negative in the dark grey region (`thin phase') and positive in the light grey region (`fat phase'). The straight lines correspond to trajectories of constant ring mass and black hole mass, i.e only the ring angular momentum varies. They have been labelled by the value of the mass fraction $m=M^{BH}/M^{BR}$.}
\label{sinalftilde}
\end{figure}

\begin{figure}[h!]
\centering\includegraphics[width=3.5in]{{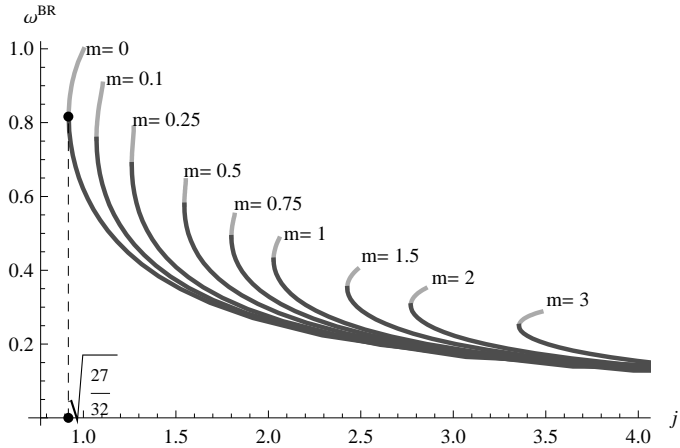}} 
\begin{picture}(0,0)(0,0)
\end{picture}
\caption{Dimensionless angular velocity of the black ring horizon $\omega^{BR}$, in terms of the reduced spin $j$, along trajectories of constant black hole and black ring mass. The behaviour in the fat (thin) phase corresponds to the light (dark) grey curves. This figure is the analogous to Fig. \ref{doublekerr}, presented in the introduction for the double Kerr system. Observe that, in the fat case, the slope of the curve decreases as $m$ increases, manifesting the increase of the moment of inertia defined in \eqref{d1}.}
\label{wvsj}
\end{figure}

In Fig. \ref{sinalftilde} the sign of $\tilde{f}(\kappa_1,\kappa_2)$ is exhibited. It shows that the correlation with the sign of $f(\kappa_1,\kappa_2)$ extends to the whole $\kappa_1,\kappa_2$ plane. This is easily understood. The change in sign for both $f(\kappa_1,\kappa_2)$  and $\tilde{f}(\kappa_1,\kappa_2)$ occurs when the denominator, which is the same for both, vanishes. In Fig. \ref{sinalftilde}, the lines of constant black hole and black ring mass are displayed. These are simply straight lines, as follows from \eqref{1par}. In Fig. \ref{wvsj} the variation of the reduced angular velocity with the reduced spin along the constant black hole and ring mass trajectories of Fig. \ref{sinalftilde} is displayed. It shows that the moment of inertia of the black ring increases, when the central black hole mass increases, in the fat phase, in close analogy with the behaviour of the moment of inertia in the double Kerr system when the distance between the two black holes is decreased.

\section{Extremal/Zero area limit}
An isolated Kerr black hole with mass $M$ has an upper bound for the angular momentum $J$ it can carry, given by
\bequ
J\le M^2 \ . \eequ
The equality is attained for the extremal Kerr solution, which has a regular degenerate horizon with finite, non-zero area. In \cite{Herdeiro:2008kq,Costa:2009wj} it was shown that, in the double Kerr system, the individual Kerr black holes may have $J>M^2$ keeping a finite, non-zero area horizon. Moreover the maximum allowed angular momentum $J_{max}$ that may be carried by the individual black holes was computed, for fixed mass, as a function of the physical distance $d$. In both the co and counter-rotating cases, $J_{max}$ increases monotonically as the distance decreases; in the counter-rotating case $J_{max}\rightarrow \infty$ as $d\rightarrow 0$; in the co-rotating case $J_{max}\rightarrow 2M^2$, as $d\rightarrow 0$. This increase in the angular momentum that may be carried by individual black holes was interpreted to be a consequence of the decrease in the angular velocity of the horizon, and therefore of dragging effects. We may now observe a similar effect in the fat phase of the black Saturn system.

An isolated black ring has no upper limit on its angular momentum for fixed mass, due to the thin phase. But isolated fat rings have both an upper and lower bound for the angular momentum they can carry, for given mass, which is expressed in terms of the reduced spin as
\bequ
\frac{27}{32}\le j^2< 1\ . \eequ
The upper bound, defined by the zero area limit, coincides with that of a single spinning Myers-Perry black hole in five dimensions; but the black hole has no lower bound \cite{ring}. Rings have a lower bound as a regularity constraint: they must rotate to be in equilibrium.

In Fig. \ref{wvsm} it is shown that, as the mass of the central black hole increases, black rings in the fat phase have smaller angular velocity along $j^2={\rm constant}$ trajectories. We have interpreted this effect as the backreaction of dragging the central black hole, which has increased the moment of inertia of the black ring - Fig. \ref{wvsj}. Moreover,  Fig. \ref{wvsm} (and Fig. \ref{wvsj}) shows that solutions with reduced spin greater than unity are allowed in both the fat and thin phase. Recall that for an isolated black ring this is only allowed in the thin phase. Observe also that as $j^2$ increases, the highest possible dimensionless angular velocity in the constant $j^2$ trajectory decreases. As for the double Kerr system, we interpret that it is the decrease in the angular velocity, due to frame dragging, that allows the ring in the fat phase to carry more angular momentum than if isolated, for the same mass.

\section{Frame dragging and the `gravitomagnetic clock effect'}
In spite of the different sign of the response function \eqref{rf1} in the fat and thin phases, which we have correlated in section \ref{correlation} with the sign of the moment of inertia, we shall now argue that the dragging effects measured by an observer at infinity \textit{always} decrease when the central black hole mass increases, for fixed ring mass and angular momentum.

We consider the following thought experiment to measure the magnitude of frame dragging effects far from a black hole. We suppose that the spacetime admits an asymptotically timelike Killing vector
$\partial/\partial t$ and a $U(1)$ action associated with a
Killing vector $\partial/\partial \psi$. Further, we assume that there
is a totally geodesic, asymptotically flat, $2+1$ dimensional submanifold, which admits a $U(1)$ action generated by $\partial/\partial \psi$. We call such a submanifold an equatorial plane. The metric on this submanifold is
\begin{equation}
g = g_{tt}(r)dt^2 + 2 g_{t \psi}(r) dt d\psi+g_{\psi\psi}(r) d\psi^2+g_{rr}(r)dr^2.
\end{equation} 
Outside the horizon, we have $g_{tt}g_{\psi\psi}-g_{t\psi}^2 <0$, and
the ergoregion corresponds to $g_{tt}>0$.

Suppose now that two satellites orbit the central body at the same radius
$r=r_0$, but in opposite directions. Associated to the central black object's rotation, and therefore to frame dragging, one satellite will orbit faster than the other. 
An observer at infinity watches the satellites and measures the
period $T_i$ of each orbit, relative to the observer's clock. We shall
see that the difference in the periods, $T_1-T_2$ is
\emph{independent} of the radius at which the satellites orbit, when
the central body is a Kerr or a (single spinning) Myers-Perry black hole, or a (single spinning) black ring;
its value depends only on the ADM mass, angular momentum and spacetime dimension.

In order to simplify matters, we introduce the following
\begin{equation}
\displaystyle G(r) = \left(\begin{matrix} g_{tt} & g_{t \psi} \\
  g_{t\psi} & g_{\psi\psi} \end{matrix} \right), \qquad T = \left(\begin{matrix} t(s) \\
  \psi(s)\end{matrix} \right).
\end{equation}
We further assume that the radial coordinate $r$ is chosen so that
$g_{rr}\equiv 1$, which simplifies the calculations although our
results will be independent of this choice. Then the Lagrangian for
geodesic motion becomes
\begin{equation}
L = \frac{1}{2} \dot{T}^{t} G(r) \dot{T} + \frac{1}{2}\dot{r}^2 \ .
\end{equation}
The Euler-Lagrange equations are
\begin{eqnarray}
\frac{d}{ds}\left(G(r) \dot{T} \right)&=&0 \ , \\
 \frac{1}{2} \dot{T}^{t} G'(r) \dot{T} &=&\ddot{r} \ .
\end{eqnarray}
Suppose then that $r=r_0$ is a circular geodesic. Clearly the first
equation implies that $\dot{T}$ is independent of $s$ (since outside
the horizon $G(r_0)$ is invertible). The second equation implies that
\begin{equation}
\dot{T}^{t} G'(r_0) \dot{T}=0 \ .
\end{equation}
Since we are dealing with a homogeneous $2 \times 2$ system now, we write $\dot{T} =
(z,1)$, where $z=dt/d\psi$ for the orbit and the equation may be written
\begin{equation}
0=\dot{T}^{t}G'(r_0)\dot{T}=g_{tt}'(r_0)z^2 + 2g_{t \psi}'(r_0) z + g_{\psi \psi}'(r_0) \ .
\end{equation}
This quadratic equation has two roots, which outside the ergoregion must have opposite signs. The positive root corresponds to the inverse angular velocity of the satellite moving in a positive sense around the central object as for a physical satellite, $dt>0$, thus if $z>0$ we must have $d\psi >0$. Similarly the negative root is the inverse angular velocity of the satellite moving in a negative sense around the central object . Let $z_\pm$ be the positive and negative roots respectively. The first satellite takes a time $t_+ = 2 \pi z_+$ to complete an orbit, while the other takes a time $t_-=-2 \pi z_-$ (the sign difference comes from the fact that it moves in the opposite sense). The difference then is
\begin{equation}
\Delta t(r_0) = 2 \pi \left(z_+ + z_- \right) = - 4 \pi\frac{ g_{t \psi}'(r_0)}{g_{tt}'(r_0)} \ . \label{drag}
\end{equation}
This is clearly a geometric quantity, so is independent of the
coordinates and we indeed see that changing the $r$ coordinate does
not alter the right hand side of our equation. Thus this continues to
be true, even when we do not assume $g_{rr} \equiv 1$. We shall provide some examples in the next subsection; but before let us note that indeed circular timelike geodesics exist, at least near infinity. 

There are two conditions in order that timelike circular orbits
exist. The condition that the roots we found above are real is 
\begin{equation}
\det{G'(r_0)}<0 \ . \label{c1}
\end{equation}
This is required so that the curve $r=r_0$ is in fact a geodesic. For
Myers-Perry, we find that this is always satisfied. We further wish the
geodesics be both timelike. This corresponds to the requirement that
\begin{equation}
\dot{T}^t G(r_0) \dot{T}<0 \label{c2} \ ,
\end{equation}
for \emph{both} the geodesics found. In the case of Myers-Perry, we require
that 
\begin{equation}
(D-1) m + 2 |a| \sqrt{(D-3) m r_0^{D-5}} - r_0^{D-3} <0 \ ,
\end{equation}
which corresponds to $r_0$ being outside the larger of the two photon
orbits (the spin of the black hole splits the co-rotating and
counter-rotating photon orbits). Note that there will be stable and unstable circular orbits, depending on the value of $r_0$.

Conditions (\ref{c1}, \ref{c2}) are stable under small perturbations to the metric functions, so clearly any asymptotically flat metric with an equatorial plane of the form considered above will admit timelike circular orbits for sufficiently large $r_0$.

\subsection{Examples}

\begin{itemize}
\item For the equatorial plane of a Kerr or a single spinning Myers-Perry black hole in $D$ spacetime dimensions, the metric coefficients are
\begin{eqnarray}
g_{tt} &=& -\left(1-\frac{2m}{r^{D-3}}\right) \ , \qquad  g_{t\psi} = -\frac{2 a m}{r^{D-3}} \ , \qquad 
 g_{\psi \psi} = a^2+r^2+\frac{2 a^2 m}{r^{D-3}}\ .
\end{eqnarray}
The formula for $\Delta t$ then gives
\begin{equation}
\Delta t(r_0) = 4 \pi a \ ,
\end{equation}
which, as advertised, is independent of $r_0$. The fact that this time difference has the same sign as $a$ means that the co-rotating satellite always takes longer to complete an orbit than the counter rotating satellite. This appears somewhat paradoxical at first glance, but observe that we are comparing orbits with different angular momentum. Actually the counter-rotating orbit must have, for the same energy $E$ and radius $r_0$, larger angular momentum (in modulus), $j_-$, than that of the co-rotating one, $j_+$, which makes the result natural.\footnote{Some observations made in \cite{Mashhoon:1998nn} about $\Delta t$ being positive seem, as such, inadequate.} Indeed, a straightforward computation shows that 
\bequ
\frac{j_{\pm}}{E}=\frac{(a^2+r_0^2)r_0^{D-3}\mp 2am\sqrt{\frac{r_0^{D-1}}{m(D-3)}}}{ar_0^{D-3}\pm(r_0^{D-3}-2m) \sqrt{\frac{r_0^{D-1}}{m(D-3)}}} \ . \eequ
Taking $a>0$, such that the black hole is rotating in the positive $\psi$ direction, $|j_-|\ge |j_+|$, with the equality attained if $a=0$.

In the four dimensional context, the proper (rather than coordinate) time difference between a circular co-rotating and counter-rotating orbit, is called the \textit{gravitomagnetic clock effect}. In \cite{CM} it was noted that, within the approximation used, such time difference was independent on the radius of the circular orbit, and suggested it could be used to measure the gravitomagnetic field of the earth. In \cite{Mashhoon:1998nn} the time difference was computed exactly in terms of coordinate time. Here we note that, in terms of coordinate time (i.e proper time of a non-rotating asymptotic observer), the $r_0$ independence is an exact result valid for Kerr/single spinning Myers-Perry black holes in any dimension. Such universality property can be expressed, using e.g.  \cite{Emparan:2008eg}, in terms of the ADM mass and angular momentum
\begin{equation}
\Delta t (r_0) =  2  (D-2)\pi\frac{J_{ADM}}{M_{ADM}} \ . \label{MP}
\end{equation}
A qualitative distinction between the four dimensional Kerr case and the six or higher dimensional Myers-Perry is that in the former case $\Delta t$ is bounded for fixed mass, whereas in the latter an arbitrarily high value can be obtained for any mass, since $J_{ADM}$ is not bounded by the mass.
\item In the coordinate system of \cite{Elvang:2007rd}, the equatorial plane outside the
black ring corresponds (after rescaling to the dimensionless
parameters) to $\rho=0$, $\bar{z} < \kappa_3$. It is convenient to
transform the radial coordinate to\footnote{This transformation can be motivated by the relation to the asymptotic spherical coordinates $(r,\theta)$
\[ \rho=\frac{L^2r^2}{2}\sin 2\theta \ , \qquad \bar{z}=\frac{r^2}{2}\cos 2\theta \ . \] Note that $r$ is a dimensionless coordinate.}
\begin{equation}
\bar{z} = 1-\frac{r^2}{2} \ ,
\end{equation}
which sends the asymptotically flat end to $r= + \infty$. We impose
the regularity conditions at the ergosphere and infinity, as in \cite{Elvang:2007rd} and finally set the constant $\bar{c}_2=0$ so that the black hole carries no angular momentum. With these choices the metric takes the form 
\begin{eqnarray}
g_{tt} &=& -\frac{(r^2-2)(r^2-2+2 \kappa_1)}{r^2(r^2-2+2\kappa_2)} \ , \\
g_{t\psi} &=& - L \frac{ 2\sqrt{2} (r^2-2+2\kappa_1)}{r^2(r^2+2\kappa_2
  -2)}\sqrt{\frac{\kappa_2 \kappa_3}{\kappa_1}} \ , \\
g_{\psi \psi} &=& L^2\left(r^2 + \frac{2(r^4 \kappa_1 +
  2((r^2+2)\kappa_1-2)\kappa_2)\kappa_3}{r^2\kappa_1(r^2+2\kappa_2-2)}\right)
\ , \\
g_{rr} &=& L^2\frac{r^2(r^2+2\kappa_2-2)}{(r^2+2\kappa_1-2)(r^2+2\kappa_3-2)} \ .
\end{eqnarray}
Where $0\leq\kappa_3<\kappa_2<\kappa_1\leq 1$. There is an ergoregion
for $r<\sqrt{2}$ and the horizon is at
$\sqrt{2(1-\kappa_3)}$. Plugging into the formula for $\Delta t$, we
find the result 
\begin{equation}
\Delta t (r_0)= 4 \pi L
\frac{r^4_0-4r^2_0(1-\kappa_1)+4(1-\kappa_2)(1-\kappa_1)}{(1-\kappa_1+\kappa_2)
  r^4_0-4(1-\kappa_1)r^2_0 +4 (1-\kappa_1)(1-\kappa_2)} \sqrt{\frac{2
    \kappa_2 \kappa_3}{\kappa_1}} \ ,
\end{equation}
which is independent of $r_0$ if $\kappa_1 = 1$, but
not otherwise. If $\kappa_1=1$ (i.e for an isolated black ring) we find
\begin{equation}
\Delta t (r_0) = 4\pi L \sqrt{\frac{2 \kappa_3}{\kappa_2}} \ .
\end{equation}
Comparing with the explicit formulae for $J_{ADM}, M_{ADM}$ given in \cite{Elvang:2007rd}, we find that
\begin{equation}
\Delta t_{\infty}\equiv \lim_{r_0\to \infty} \Delta t = 6 \pi \frac{J_{ADM}}{M_{ADM}} = 2(D-2) 
\pi\frac{J_{ADM}}{M_{ADM}} \ .
\end{equation}
Note that this formula agrees with that of Myers-Perry black holes \eqref{MP}. Thus, by doing this thought experiment, the asymptotic observer, at $r_0=\infty$, could not distinguish between a Myers-Perry black hole, a black ring or a black Saturn with the same conserved quantities.
\end{itemize}

\subsection{A measure of frame dragging at infinity}
Using equation (\ref{drag}), and letting $r_0 \to \infty$, we see that the time difference $\Delta t$, which is then a measure of frame dragging at infinity, is given, for a general stationary, asymptotically flat spacetime admitting a $U(1)$ symmetry, in terms of the ADM mass and angular momentum by
\begin{equation}
\Delta t_\infty =  2 (D-2)\pi \frac{J_{ADM}}{M_{ADM}} \ . 
\end{equation}
We interpret this quantity as giving us a rough idea of the degree to which the central objects are dragging the spacetime near infinity. Applying to the black Saturn system we discussed, for which $M_{ADM}=M^{BH}+M^{BR}$, and $J_{ADM}=J^{BR}$, we see that, in our thought experiment, increasing the mass of the central black hole, for fixed ring mass and angular momentum, always \emph{decreases} the amount of frame dragging at infinity, in both the fat and the thin phases.

\section{Black Saturn with $J^{BH}\neq 0$}
\label{sectionj}
\begin{figure}[h!]
\centering\includegraphics[width=3.5in]{{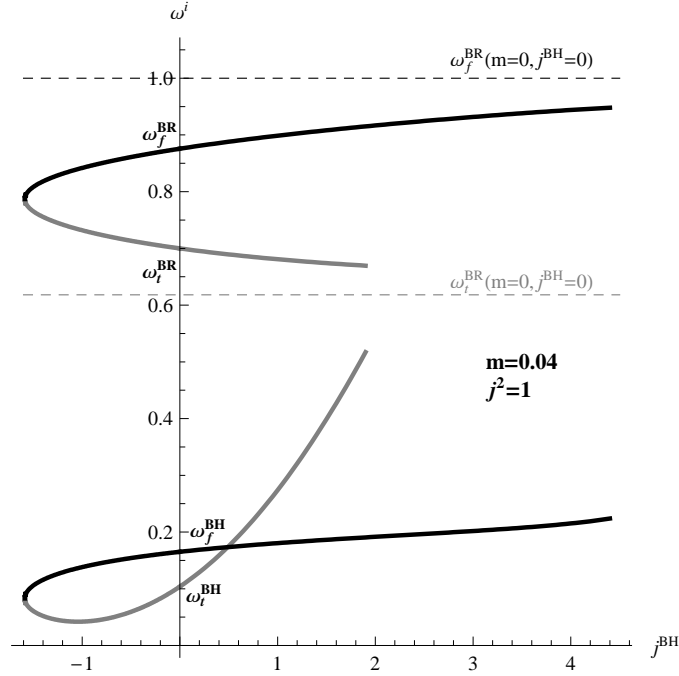}} 
\begin{picture}(0,0)(0,0)
\end{picture}
\caption{Dimensionless angular velocity of the black ring and black hole horizons, in the fat, $\omega^{BR}_f$ and $\omega^{BH}_f$, and thin,  $\omega^{BR}_t$ and $\omega^{BH}_t$, phases, in terms of the reduced angular momentum of the black hole. The angular momentum of the black hole is turned on with either the same sign (co-rotating case, $j^{BH}>0$) or opposite sign (counter-rotating case, $j^{BH}<0$) as the black ring angular momentum. The dashed lines correspond to the reduced angular velocity of an isolated fat (top line) and thin (bottom line) black ring with with $j^2\simeq 1$.}
\label{jnotzero}
\end{figure}

In the double Kerr system, the angular velocity of the black holes decreases, as the physical distance decreases, even when the two black holes are co-rotating. An analogous situation can be observed in the black Saturn system considering $J^{BH}\neq 0$. To allow comparison with an isolated (fat) black ring, we take the reduced spin for the ring to be $j^2=1$. From Fig. \ref{wvsm} we see that this constrains the relative mass of the system to be smaller than $m\simeq 0.05$. We take $m=0.04$. Since the intrinsic angular momentum of a single spinning Myers-Perry black hole in $D=5$ is bounded by the mass, there will be an upper bound for the reduced spin of the black hole. The analysis of the dimensionless angular velocities of the two black objects is displayed in Fig. \ref{jnotzero}.

Various features can be observed from Fig.  \ref{jnotzero}:
\begin{itemize}
\item{Fat phase:} turning on $j^{BH}>0$, the angular velocity of both black objects increases. For the black ring this is due to effect i). For the black hole this is the statement that it has positive moment of inertia. Notice, however, that the angular velocity of the black ring always remains smaller than that of an isolated ring with the same conserved charges. This is analogous to the behaviour observed in the double Kerr system for the co-rotating case, as advertised. Such a result shows that for rings in the fat phase, as for the cases discussed of the double Kerr system, effect ii) dominants over effect i), since in the co-rotating case the former (latter) effect is expected to produce a speed up (slow down) of the angular velocity of the ring horizon. Analogously,  turning on $j^{BH}<0$, the angular velocity of both black objects decreases. 
\item{Thin phase:} turning on $j^{BH}$, for either sign, the angular velocity of the black ring has the opposite behaviour to that observed in the fat phase. The angular velocity of the black hole always increases when $j^{BH}>0$; for $j^{BH}<0$, it initially decreases, as expected, but then starts increasing  for $j^{BH}\simeq -1$. Results in the thin phase are harder to interpret. We have verified that the proper distance along the equatorial plane between the two black objects increases, as $j^{BH}$ increases. This accommodates, geometrically, the fact that $\omega^{BR}_t$ decreases as $j^{BH}$ increases. Thus, for the smallest possible values of $j^{BH}$, the two objects will be close enough for the ring to exerts a strong rotational dragging (effect i)) on the black hole, explaining the inversion observed in the behaviour of $\omega_t^{BH}$ for $j^{BH}<0$.
\end{itemize}

\section{Final remarks}
Recent studies of the double Kerr system  \cite{Herdeiro:2008kq,Costa:2009wj} have suggested a number of physical effects related to rotational dragging in a system with multiple black holes, namely:
\begin{description}
\item[$a)$] Dragging a heavy object (such as a black hole) backreacts on the dragging source (another black hole) by decreasing its angular velocity, as compared to the angular velocity of the same (in terms of conserved quantities), but isolated, source.
\item[$b)$] The decrease in the angular velocity allows, for fixed mass, a higher angular momentum than if the black hole were isolated. Thus, extremal black holes have $J>M^2$ in this system.
\end{description}
The aim of this paper was to confirm and understand these effects in a different multi-black hole solution with a clear physical advantage: the balanced black Saturn is a regular (on and outside an event horizon) solution of the vacuum Einstein's equations and consequently has no further parameters besides the conserved charges of each black object. Thus, the confirmation of the effects in this system establishes that they are not an artefact of the strut (conical singularity) present in the double Kerr system. Let us summarise what has been learned about the effects just mentioned:  
\begin{description}
\item[$a')$] Studying a black Saturn with a zero angular momentum central black hole, shows that the backreaction on the black ring of dragging the central black hole may decrease or increase the angular velocity of the ring. For small black hole mass the two phases occur, respectively, for fat or thin rings. In general, we have dubbed the two black Saturn phases as fat or thin phases.\footnote{The existence of two phases in the black Saturn system with analogous properties to those of fat/thin rings was discussed in the original work \cite{Elvang:2007rd}, wherein the manifestations of effect i) discussed in the introduction were also observed and characterised.} We have correlated these phases with the sign of another response function which is interpreted as an inverse moment of inertia. In the fat (thin) phase, the moment of inertia is positive (negative). In the double Kerr system the analogous moment of inertia is positive, and therefore the system should be compared with the fat phase. Thus the black Saturn results confirms effect $a)$ above, but simultaneously show that there are gravitational system with exactly the opposite behaviour. 
\item[$b')$] Fat black rings have an upper bound for the angular momentum for fixed mass. As for the black holes in the double Kerr system, the decrease in the angular velocity in the fat phase of the black Saturn system allows, for fixed ring mass, a higher angular momentum than if the black ring were isolated. Thus, in the fat phase, black rings may have $j^2>1$ in this system.

\end{description}

Studying a black Saturn having a central black hole \textit{with} intrinsic angular momentum allowed a comparison between effect i) and ii) on the black ring. For the fat phase, the result supports the interpretation that effect ii)  described in the introduction (the backreaction of dragging the central black hole) is dominant over effect i) (the dragging exerted by the black hole). Again this finds a close parallelism in the double Kerr system: the maximum angular velocity for each black hole in this system is smaller than the maximum angular velocity of the same black hole in an isolated system, even in the co-rotating case.

In spite of the counter intuitive behaviour of thin rings, that increase their angular velocity when dragging a heavier black hole, we pointed out that a physical measure of frame dragging given by the `gravitomagnetic clock effect' establishes that frame dragging, as measured by an observer at infinity, always decreases in the black Saturn system when the central black hole mass is increased for fixed ring mass and angular momentum. In a nut shell, `dragging always costs'.

Finally, the interaction effects between black holes discussed in this paper, could be studied in other exact solutions with multiple black holes, namely the di-ring solution \cite{Evslin:2007fv}.

\section*{Acknowledgements}
We would like to thank L. F. Costa for discussions. C.H. would like to thank the hospitality of D.A.M.T.P., University of
Cambridge, where part of this work was done. C.H. is supported by a ``Ci\^encia 2007" research contract. C.R. is funded by FCT through grant SFRH/BD/18502/2004. C.M.W. thanks Queens' College, Cambridge for a Research Fellowship. This work as been further supported by the FCT grant CERN/FP/83508/2008.

\end{document}